# Teaching and learning special relativity theory in secondary and lower undergraduate education: A literature review

Paul Alstein®,[*] Kim Krijtenburg-Lewerissa®, and Wouter R. van Joolingen®
*Freudenthal Institute, Utrecht University, P.O. Box 85170, 3508 AD Utrecht, Netherlands*

This review presents an overview and analysis of the body of research on special relativity theory (SRT) education at the secondary and lower undergraduate level. There is currently a growing international interest in implementing SRT in pre-university education as an introduction to modern physics. For this reason, insights into learning opportunities and challenges in SRT education are needed. The field of research in SRT education is still at an early stage, especially at the level of secondary education, and there is a shortage of empirical evaluation of learning outcomes. In order to guide future research directions, there is a need for an overview and synthesis of the results reported so far. We have selected 40 articles and categorized them according to reported learning difficulties, teaching approaches, and research tools. Analysis shows that students at all educational levels experience learning difficulties with the use of frames of reference, the postulates of SRT, and relativistic effects. In the reported teaching sequences, instructional materials, and learning activities, these difficulties are approached from different angles. Some teaching approaches focus on thought experiments to express conceptual features of SRT, while others use virtual environments to provide realistic visualization of relativistic effects. From the reported teaching approaches, three learning objectives can be identified: to foster conceptual understanding, to foster understanding of the history and philosophy of science, and to gain motivation and confidence toward SRT and physics in general. In order to quantitatively compare learning outcomes of different teaching strategies, a more thorough evaluation of assessment tools is required.

## I. INTRODUCTION

In recent decades, physics education has aimed to promote understanding of the nature of science in addition to domain-specific content [1,2]. An important aspect of the nature of science is that well-established concepts in a scientific discipline can be changed radically by the introduction of a new theory. A prime example of this process is provided by special relativity theory (SRT), in which the classical notion of space and time as separate, absolute entities is replaced by the notion of spacetime [3], which has become fundamental to the development of modern physics. It is increasingly recognized that learning about the conceptual and epistemological leap between classical and modern physics may support a better understanding of scientific reasoning and improve students' motivation and attitude toward physics [4,5].

While traditionally SRT is introduced at the lower undergraduate level, there is a growing international interest in including SRT into the secondary education curriculum. Currently, an introduction to SRT is included in the secondary school curriculum in Netherlands [6], Norway [5], Germany [7], Argentina [8], Australia [9], and South Korea [10]. The international Einsteinian Physics Education Research collaboration is currently investigating how to implement the learning domain of modern physics in primary education [11,12].

One of the main aims of SRT education is to illuminate the conceptual leap in our understanding of space, time, and motion. This objective favors a conceptual teaching approach, in which SRT is conceptualized through the use of thought experiments (TEs), rather than a mathematical approach that emphasizes the formalism of Lorentz transformation or the geometry of spacetime [13,14]. At the level of secondary education, moreover, students have not yet mastered the mathematical skills required for the fourvector formalism. In order to foster conceptual understanding, students are required to explore their existing conceptions and beliefs and to evaluate them after instruction [4,15]. A variety of student conceptions regarding SRT has been identified in prior research, and it is argued that these form a source of robust learning difficulties [13,16].

---

[*]p.alstein@uu.nl

A number of teaching approaches have been developed to overcome these difficulties and to advance students' motivation and confidence, although this field of research is still at an early stage, and there is a shortage of empirical evaluation of learning outcomes. Consequently, there is no clear consensus on research-informed guidelines for the design of teaching approaches and teaching materials, nor on the research tools used to investigate students' reasoning and understanding. Moreover, the majority of studies on learning difficulties and teaching approaches focus on undergraduate education, while learning opportunities and challenges at the secondary school level have received much less attention. To meet the growing demand for insights in teaching and learning SRT, there is currently a need for a general overview and synthesis of the results reported so far.

This review study summarizes and analyzes the body of research in the field of SRT education at the secondary and lower undergraduate level, focusing on learning difficulties, teaching approaches, and research tools. We limit our review to a conceptual approach to relativistic kinematics, including prerequisite concepts from classical mechanics, specifically the use of frames of reference in Galilean relativity. More specifically, we will address the following research questions:

1. Which learning difficulties in SRT and Galilean relativity education have been reported at the secondary and lower undergraduate level?
2. Which teaching approaches have been developed for SRT education at the secondary and lower undergraduate level?
3. Which research tools have been used to probe students' conceptual understanding of SRT at the secondary and lower undergraduate level?

## II. METHOD

To answer the research questions, three online databases were searched: ERIC, Scopus, and Web of Science. The following search terms were used to find relevant articles from reviewed journals: ("special relativity" OR "Galilean relativity") AND ("education" OR "teaching" OR "student"). In Scopus, the search results where limited to articles in the subject area of physics and astronomy. This search resulted in 415 articles, 117 of which were identified as duplicates and 8 of which were written languages other than English. Two articles were added by means of cross reference, as they use different terms in the title and abstract [17,18].

Subsequently, the following criteria were used to filter the search results: (i) the article addresses conceptual understanding of SRT or Galilean relativity without presupposing an understanding of the formalism of Lorentz transformation or spacetime geometry, (ii) the article applies to secondary or lower undergraduate education, where SRT is introduced for the first time, (iii) the article

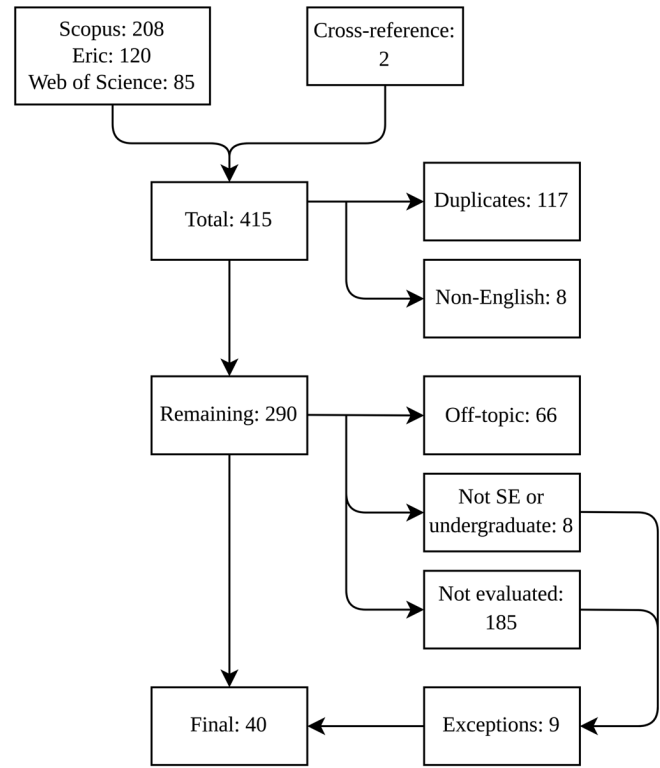

FIG. 1. A flow chart of the selection process, indicating the number of selected articles for each step. The abbreviation SE stands for secondary education.

uses research-based tools to identify learning difficulties or evaluate teaching approaches.

A number of exceptions were made for articles that do not match one of these criteria. Exceptions to criterion 2 were made for a pivotal study by Villani and Pacca that focuses on graduate education [19] and two articles that report relevant insights at the level of preservice teacher education [20,21]. Exceptions to criterion 3 were made for articles that report novel multimedia tools that have not yet been empirically evaluated [22–26,52]. A flow chart of the selection process is shown in Fig. 1.

The selection process resulted in a total of 40 articles. The following sections present an analysis of the reported learning difficulties, teaching approaches, and research tools.

## III. LEARNING DIFFICULTIES

This section addresses the first research question: "Which learning difficulties in SRT and Galilean relativity education have been reported at the secondary and lower undergraduate level?" In order to identify learning difficulties, it is useful to describe students' learning progression in a typical introductory course on SRT. The learning progression presented here is based on an analysis of college physics textbooks by Dimitriadi and Halkia [13]. Any description of relative motion is grounded in the concept of frames of reference, in which observers preform measurements to determine spatial and temporal coordinates of events. Hence, solid understanding of the transformation of measurement outcomes between frames of reference may be considered prerequisite for an introduction to SRT. Once students have mastered the ability to switch between frames of reference, it is possible for them to recognize the profoundness of the postulate of special relativity and the light postulate. Because the combination of these postulates is incompatible with Galilean relativity, this step in the learning progression represents a conceptual leap with respect to classical mechanics. Once students have gained an understanding of the postulates of SRT, they can proceed to deduce relativistic effects from the postulates, following Einstein's approach in his original 1905 paper [27]. A typical introductory course on SRT at the secondary level addresses the relativity of simultaneity, time dilation, length contraction, and relativistic velocity addition, while at the undergraduate level mass-energy relation is commonly also addressed. The cumulative nature of the learning progression described above indicates that learning difficulties are also cumulative. Specifically, understanding relativistic effects presupposes a solid understanding of the postulates, which in turn requires solid understanding of frames of reference. A schematic representation of the learning progression is presented in Fig. 2. To answer the research question, the selected articles were searched for learning difficulties in three categories: (A) frames of reference, (B) postulates of SRT, and (C) relativistic effects. An overview of the main learning difficulties is presented in Table I.

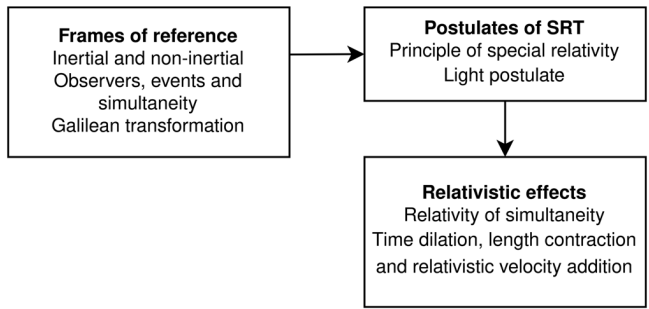

FIG. 2. A schematic representation of students' learning progression in a typical introductory course on SRT.

TABLE I. An overview of learning difficulties organized into three main categories and eight subcategories.

| (Sub)category | Student conceptions |
|---|---|
| Frames of reference | |
| General | Treating frames of reference as concrete objects, fixed to bodies [21,28] |
| | Associating particular events with particular frames of reference [28] |
| | Allowing value judgement of motion as "real" or "apparent" [13,28] |
| Inertial and noninertial | Asserting the inertial or noninertial nature of a frame of reference intuitively [29] |
| | Regarding the inertial or noninertial nature of a frame of reference as a relative property [29] |
| | Regarding pseudoforces as apparent or "nonexisting" [29] |
| Events, observers, and simultaneity | Associating the time of an event with the time at which an observer receives a signal from the event [13,16] |
| | Treating observers at relative rest in different positions as being in separate frames of reference [16,30] |
| | Treating observers at the same position as being in the same frame of reference, regardless of their relative motion [13,16,30] |
| Galilean transformation | Regarding velocity as an intrinsic property of a body [17] |
| | Allowing violation of the invariance of time intervals in the nonrelativistic limit [29] |
| | Granting the invariance of distance between two events regardless of whether they are simultaneous [29] |
| Postulates of SRT | |
| Principle of special relativity | Ascribing the equivalence of inertial frames of reference to the nonexistence of pseudoforces [31] |
| | Considering a collection of bodies as being in "true rest" when all parts of the collection move together [31] |
| | Relating the equivalence of inertial frames of reference to their kinematic reciprocity [32] |
| | Regarding the principle of relativity as a necessary condition for the transformation of physical quantities [32,33] |
| Light postulate | Considering that the "true" speed of light can be observed only in the rest frame of the light source [19] |
| | Applying a uniform speed of light relative either to the light source, the background space or a mixture of both [6] |
| Relativistic effects | |
| Relativity of simultaneity | Regarding the relativity of simultaneity as an artifact of signal travel time [16] |
| Time dilation, length contraction, and relativistic velocity addition | Considering time dilation and length contraction as unilateral phenomena, occurring only in the "moving" frame of reference [20,34] |
| | Granting that length contraction occurs in all dimensions, even when the motion is one dimensional [20,21,33] |

### A. Frames of reference

At the heart of any description of relative motion lies the conceptualization of frames of reference. In a frame of reference, events are given spatial and temporal coordinates, as measured by an observer in relative rest to the frame of reference, by means of measuring rods and synchronized clocks. Learning difficulties may arise as a result of existing conceptions with regard to the roles of events and observers and the consequences of invariance under transformation. The present section will present student conceptions found in the selected literature with regard to frames of reference in general, inertial and noninertial frames of reference, events, observers and simultaneity, and Galilean transformation.

Students' understanding of frames of reference was first investigated by Panse et al. [28], who administered a diagnostic forced-option test to undergraduate students ($N = 111$). In the wide range of student conceptions found, some patterns could be traced. First, Panse et al. showed that students tend to treat frames of reference as being fixed to concrete objects, in the sense that they are localized and physically extended to the dimensions of that object. This conception was also found in a similar study by Tanel [21], who additionally concluded that students do not take events into account that lie outside of the frame of reference's supposed physical extend. Second, students tend to regard particular events as belonging to a particular frame of reference, and fail to recognize that events can be observed regardless of the chosen frame of reference. Third, it was found that students use frames of reference to allow value judgment of motion as real or apparent. For example, students tend to grant a special status to specific frames of reference, most naturally the frame of reference of the Earth. As pointed out by Dimitriadi and Halkia [13], the notion of a preferred frame of reference may correspond to a belief in an absolute space.

#### 1. Inertial and noninertial frames of reference

An inertial frame of reference is a frame of reference in which Newton's first law of motion holds. Hence, it is possible to assert the inertial or noninertial nature of a frame of reference intrinsically by verifying that an object with no external force acting upon it remains at rest or moves at a constant velocity. By this definition, it follows that any frame of reference that moves uniformly with respect to an inertial frame of reference, must also be an inertial frame of reference.

Students' understanding of inertial and noninertial frames of references was first investigated by Ramadas et al. [29], by administering a forced-option test to undergraduate physics students ($N = 77$). It was found that students often fail to recognize the intrinsic nature of inertial frames of references as described above. Instead, students assert the inertial or noninertial nature of a frame of reference intuitively by considering it inertial when it can be regarded as "stationary." For example, students may regard the frame of reference of an accelerating rocket as inertial because it is intuitively considered as stationary by an observer inside the rocket. Second, the intrinsic nature of inertial frames of reference is ignored by regarding it as a relative property between two frames of reference. This notion may lead to the view that all frames of reference are equivalent, whether they are inertial or not. Finally, pseudoforces are treated by students as apparent or nonexisting, while in fact they are frame-dependent quantities. This conception reflects the value judgment of frames of reference found by Panse et al. [28], as mentioned in the previous section.

#### 2. Events, observers, and simultaneity

The measurement procedure for observing events in a frame of reference, as described above, ensures that observers in the same frame of reference must agree on the outcome of a measurement, regardless of the position of the observer. This implies that time measurements are not dependent on the travel time of light signals from the event toward the observer, as an observer can simply refer to a clock in its vicinity that is synchronized with a clock located at the event.

Students' understanding of time and simultaneity was investigated by Scherr et al. [16], by analysis of physics and nonphysics students' responses to a written questionnaire ($N = 800$). It was found that students tend to associate the time of an event with the time at which a light signal from the event is received by the observer. This leads to the incorrect notion that events are simultaneous if an observer receives signals from two events simultaneously. Students who hold this conception of time measurement will consequently tend to treat observers at different positions in the same frame of reference as disagreeing with the time coordinate of the event. Conversely, Scherr et al. showed that students tend to treat observers at the same position as being in the same frame of reference, regardless of their relative motion. Generally, these student conceptions indicate a view where each observer constitutes a distinct frame of reference, where outcomes of measurements are dependent on the position of the observer at any given time. This view was replicated in similar studies by de Hosson et al. [30] and Dimitriadi and Halkia [13].

### 3. Galilean transformation

Switching between observations in one frame of reference to another, the spatial and temporal coordinates of an event are transformed into coordinates associated with the new frame of reference, as dictated by the relative motion between the two frames of reference. In the nonrelativistic limit, coordinates of an event in two inertial frames are related by Galilean transformation, which reflect the spatial and temporal symmetries of Newtonian space and time. In particular, the distance between non-simultaneous events is dependent on the relative motion of the two inertial frames of reference, while the time interval between any two events is preserved. Consequently, velocity is a relative quantity in Galilean relativity.

Students' understanding of transformation of physical quantities between inertial frames of references in Galilean relativity was first studied by Saltiel and Malgrange [17] by means of a set of open-ended questions administered to undergraduate physics students ($N = 700$). The results illustrate a "spontaneous model" of reasoning about kinematics, in which velocity is viewed as an intrinsic property of a body and trajectories are described independently of the chosen frame of reference.

This outcome was replicated in a study by Ramadas et al. [35], where a forced-option test was administered to undergraduate physics students ($N = 102$). In addition, Ramadas et al. showed that students do not consider violation of the invariance of time intervals as problematic, and grant invariance of distance between two events regardless of whether they are simultaneous. These results generally indicate that students have a strongly held belief in motion as an absolute entity.

### B. Postulates of special relativity

In Einstein's original 1905 paper [27], SRT is derived from two postulates: the principle of special relativity and the light postulate. The present section presents learning difficulties reported in the selected articles with regard to the postulates.

### 1. Principle of special relativity

The principle of special relativity states that observers in any two inertial frames of reference must agree on the laws of physics. An important consequence of this principle is the equivalence of inertial frames of reference, making it impossible to determine whether any frame of reference is moving or resting. This principle is essentially equivalent to Galilei's principle of relativity, but including electrodynamics and optics in addition to mechanics.

A wide variety of student difficulties in understanding and applying the principle of relativity and its profound consequences on the nature of motion has been reported. Pietrocola and Zylbersztajn [31] investigated the extent to which students spontaneously employ the principle of relativity. In their study, undergraduate physics students ($N = 21$) were presented with a variety of physical phenomena (e.g., a swinging pendulum, boiling water or electrical interaction) and asked whether a change in velocity would affect these phenomena. It was found that students do not spontaneously use the principle of relativity as a heuristic tool and, moreover, do not consider situations in which the principle of relativity is violated as problematic. Rather, students ascribe the equivalence of laws among inertial frames of reference to the nonexistence of pseudoforces in inertial frames of reference. Moreover, students consider a collection of bodies as being in true rest when all parts of the collection move together. This result indicates that some students hold a pre-Newtonian view of motion, where the "true motion" of a body is determined with respect to its immediate surrounding bodies.

Similar views were reported in a study by Bandyopadhyay [32], in which different meanings that students ascribe to the principle of relativity were investigated by analysis of responses of undergraduate physics students to a written questionnaire ($N = 20$). It was found that students relate the principle of relativity to the fact that observers in inertial frames of reference agree on their relative velocity. This kinematic reciprocity, however, holds in any two frames of reference, regardless of whether they are inertial or not. Moreover, Bandyopadhyay found that students tend to regard the principle of relativity as a necessary condition for any transformation of physical quantities. For example, one student suggested that the principle of relativity naturally implies Galilean transformation. This difficulty to delineate between the transformation of physical quantities and invariance of laws was replicated in a study by Ozcan [33].

### 2. Light postulate

The light postulate states that "light is always propagated in empty space with a definite velocity $c$ which is independent of the state of motion of the emitting body" [27]. This postulate, combined with the principle of relativity, enabled Einstein to put forward his claim against the existence of a luminiferous ether, and paved the way for his relational notion of space and time. In the selected articles, learning difficulties regarding the light postulate have received little attention compared to learning difficulties regarding the principle of relativity.

Students' preinstructional ideas about the speed of light were first investigated by Villani and Pacca [19]. By analyzing answers of graduate students ($N = 30$) to two written questions about relativistic situations, the authors found students' views on light propagation to be highly compatible with the "spontaneous model" of kinematics as described by Saltiel and Malgrange [17]. In this view, there exists a true value of the speed of light, although it is observed only in the rest frame of the light source.

Kamphorst et al. [6] further investigated preinstructional reasoning with the speed of light by analysis of clinical interviews with secondary school students ($N = 20$). In their study, event diagrams were used to explicate students' ideas of light propagation. It was found that students consistently apply a uniform speed of light in either the frame of reference of the light source, the frame of reference of the event diagram, or a mixture of both across different tasks. It is argued that students are aware that light would travel at different speeds relative to observers that are not in the preferred frame of reference. However, they do not spontaneously find this problematic.

### C. Relativistic effects

Einstein [27] showed that respecting both the principle of relativity and the light postulate leads to a radically different notion of space, time, and motion, giving rise to relativistic effects such as the relativity of simultaneity, time dilation, length contraction, and relativistic velocity addition. In the present section, learning difficulties with regard to relativistic effects are presented.

#### 1. Relativity of simultaneity

In his 1905 article [27], Einstein illuminated the consequences of the two postulates by means of a simple TE. Consider a rigid rod with synchronized clocks located at each end, positioned at the origin of an inertial frame of reference $S'$, which is in motion relative to a stationary inertial frame $S$. Applying both the principle of relativity and the light postulate, an observer in $S'$ would find that the two clocks are indeed synchronized, while an observer in $S$ would find that they are not.

The body of student conceptions found by Scherr et al. [16] with regard to time and simultaneity, result in a number of learning difficulties with regard to the relativity of simultaneity. Specifically, the tendency to associate the time of an event with the time at which an observer receives a signal from the event leads to the view that the relativity of simultaneity is an artifact of signal travel time. In this view, simultaneity is regarded as independent of relative motion between the event and the observer, as it is in the nonrelativistic limit. This outcome indicates that students have difficulties in applying Einstein's procedure for measuring time and time ordering by means of synchronized clocks.

#### 2. Time dilation, length contraction, and relativistic velocity addition

Einstein continued his 1905 TE [27] by considering how the synchronized clocks on the far ends of the rod can be used to measure the time interval between two events at the far ends of the rod. It follows from the postulates that an observer in $S'$ must disagree with the observer in $S$ on the time interval between these events. When measuring the length of the rod by simultaneously ascertaining the positions of its far ends, the observers must also disagree on the length of the rod. As the relative velocity between the events and observer approaches the speed of light, the effects of time dilation and length contraction approach infinity and zero, respectively, hence SRT requires that the maximum speed in nature is set to the speed of light.

Student conceptions of time dilation and length contraction were specified in a study by Selçuk [20], by analysis of written questionnaires and interview sessions with preservice teachers ($N = 185$). It was found that students fail to recognize the symmetry of time dilation and length contraction, considering instead that they are unilateral phenomena that occur only in the moving frame of reference. Moreover, it was indicated by Selçuk that students believe that the shortening of length should occur in all dimensions, while in fact it is restricted to the direction of motion. These views were replicated in similar studies by Aslanides and Savage [34], Tanel [21], and Ozcan [33].

It was shown by Dimitriadi and Halkia [13] that students have difficulties in realizing that the maximum speed is an intrinsic property of nature. Notably, this is the only reported learning difficulty regarding relativistic velocity addition.

### D. Synthesis

The selected articles show that students experience learning difficulties with the use of frames of reference, the postulates of SRT and relativistic effects. As argued by Scherr [16,36], even graduate students who have already studied SRT at the undergraduate level often lack understanding of underlying concepts and principles. This result is replicated in studies on learning difficulties in secondary education [6,13] and preservice teacher education [20,21,33]. It may be concluded that the conceptual leap required to understand SRT presents a challenge for students at all levels of education.

Our analysis shows that difficulties in understanding the use of frames of reference, and switching from one frame of reference to another may be caused by existing conceptions and beliefs that are rooted in everyday experience. For example, students tend to regard frames of reference as having a physical extension and regard events and observers as belonging to a particular frame of reference. This notion allows students to view motion as true or apparent, depending on the chosen frame of reference, with a special status often granted to the Earth's frame of reference. This view of motion supports the widespread belief that velocity and trajectory are intrinsic properties of a body. Finally, students tend to apply a procedure of time measurement that is based on receiving light signals from events, disregarding the equivalence of observers in the same frame of reference.

These conceptions are clearly at odds with the principle of relativity, which rejects any inequivalence between inertial frames of reference. Some researchers have argued that a stronger emphasis on Galilean transformation would help students to appreciate the profound consequences of the principle of relativity [16,36]. Because Galilean transformation does not involve any counterintuitive concepts, this topic could be learned at an early age, as a means to prepare students for an introduction to SRT in a later phase of their education. The introduction to the light postulate and its incompatibility with Galilean transformation should then lead students to recognize the necessity of a new theory of relative motion.

## IV. TEACHING APPROACHES

This section addresses the second research question: "Which teaching approaches have been developed for SRT education at the secondary and lower undergraduate level?" In the selected articles, a variety of instructional approaches, teaching materials, multimedia applications and student activities is reported, focusing on conceptual understanding of SRT as well as motivational and attitudinal objectives. Because SRT education is represented by a relatively young and growing field of research, not all teaching approaches have been thoroughly evaluated for learning outcomes or gained motivation. Consequently, this section does not seek to provide a quantitative comparison of teaching approaches that would serve as a basis to determine best practices for teachers. Rather, this section serves to illustrate the various teaching approaches reported so far and presents an analysis of the learning objectives at which they aim.

The reported teaching approaches are grouped into four categories: (A) focus on TEs, (B) historical and philosophical contextualization, (C) multimedia support, and (D) student activities. The first two categories mainly describe teaching sequences and instructional materials, while the last two categories describe specific learning tools and student activities.

### A. Focus on thought experiments

The use of TEs is central to the conceptual approach to SRT, as it allows teachers and students to express underlying principles, mechanisms, and assumptions in situations that relate to everyday experience. In this regard, TEs represent an important characteristic of scientific reasoning [18,37]. This section presents a discussion of teaching approaches that rely heavily on the use of TEs. A detailed discussion of the reported TEs is presented in Sec. V.

Scherr *et al.* [36] reported the development of instructional tutorials intended to improve students' understanding of the concepts of time, the relativity of simultaneity and the roles of observers in inertial frames of reference. The design of the tutorials was based on the authors' previous analysis of learning difficulties [16], as described in Secs. III A 2 and III C 1. After an introductory instruction on SRT, two tutorials were handed out: one focusing on measurements of time in a single frame of reference and one focusing on measurements of simultaneity across multiple frames of reference. Each tutorial consists of three steps: first, a pretest was given, designed to elicit student conceptions; second, students were led to recognize a discrepancy between their conceptions and the concepts according to the new theory; and third, students were guided through the reasoning necessary to resolve any inconsistencies. An example of a TE scenario used to express the process of time measurement involved two distant erupting volcanoes and an observer on the ground who measures the simultaneity or time ordering of the eruptions. To foster meaningful learning, it was argued, the confrontation and resolution between student conceptions and the new concepts must be carried out by the students themselves, rather than by the teachers. Analysis of three pre-post tests ($N = 300$) and excerpts from taped interviews and classroom interactions showed that the materials help to confront incompatibility of deeply held student conceptions about simultaneity and improve students' ability to recognize and resolve some of the classic paradoxes of SRT.

Dimitriadi and Halkia [13] reported a teaching and learning sequence for the upper levels of secondary education, aimed at conceptual understanding of the postulates of SRT and relativistic effects. Based on analysis of physics college textbooks, a review of relevant literature, and a pilot study, a teaching and learning sequence was designed. The sequence consisted of 5 lessons, focusing on the principle of relativity, the light postulate, the relativity of simultaneity, time dilation, and length contraction. In order to make the material suitable for upper secondary school students, mathematical formalism and complicated terms, such as "inertial frame of reference," "length contraction," and "time dilation" were avoided, as well as any mention of the luminiferous ether and the Michelson and Morley experiment. Instead, the authors made use of TEs, simple short stories that connect to everyday experience and passages from popularized science books. Analysis of clinical interviews and a pre-post open-ended questionnaire ($N = 40$) showed that upper secondary school students are able to cope with the basic ideas of SRT. However, there were some difficulties caused by students' deeply held belief in absolute motion and the supposed dependence on the observer's position in its frame of reference. The authors concluded that students experience more difficulties with applying the principle of relativity than with applying the light postulate. Moreover, the authors concluded that students seemed to have no difficulty in dealing with problems that do not relate to everyday experience.

Velentzas and Halkia [18] investigated how TEs can be used as didactical tools in teaching SRT and general relativity in upper secondary education. Two TEs were selected from an earlier analysis of textbooks and popularized physics books [38]: one describing the relativity of simultaneity, time dilation, and length contraction ("Einstein's train"), and one describing the principle of equivalence in general relativity ("Einstein's elevator"). These TEs were presented to students in semistructured interviews ($N = 40$), and the students were subsequently asked to predict their outcomes. Analysis of tape recordings of the interviews and three open-ended post-test questions showed that TEs represent useful teaching and learning tools. The authors argued that the narrative form and minimum of mathematical formalism of TEs increased students' motivation and helped students to approach abstract concepts and principles that do not relate to everyday experience. Some students found it difficult to accept the counterintuitive consequences of SRT. However, this barrier could be overcome by realizing the difference between reasoning about everyday situations and scientific reasoning.

Otero et al. [39] described a didactic sequence for the upper level of secondary education, aimed at investigating the conceptualization of the basic aspects of SRT. The didactic sequence focused on eight TEs, each complemented with a set of exercises. These TEs were grouped into three parts: Galilean transformation, the postulates of SRT, and the relativity of simultaneity, time dilation, and length contraction. A typical example of a TE scenario involved a wagon that is moving with respect to an external observer, with an internal observer located in the middle of the vehicle, and shooting either rubber bullets or light rays to opposite sides of the wagon. Students were then asked to describe the trajectories of the rubber bullets or light rays from the perspective of both observers. Analysis of student responses to the exercises ($N = 43$) showed that student difficulties were mostly related to unsuccessful conceptualization of the principle of relativity. Contrary to the conclusions of Dimitriadi and Halkia [13], the authors concluded that the light postulate was more easily accepted.

A follow-up study [8] investigated the conceptualization of the relativity of simultaneity, using the same research methods as the original study. Analysis of student responses to the exercises ($N = 128$) showed that the proposed TEs produce the necessary awareness of important aspects of Galilean transformation, paving the way for the conceptualization of the relativity of simultaneity.

### B. Historical and philosophical contextualization

The objective of teaching students about the nature of science has drawn attention to the benefits of introducing physics from a historical and philosophical point of view [40,41]. Not only does explicit reference to debates from the history and philosophy of science (HPS) promote a better understanding of the nature of science as a human and cultural achievement, it also facilitates a connection between the development of individual thinking and the development of scientific thought [42]. Transforming historical arguments into teaching practices may help to reveal the urgency of a new theory and, conversely, to locate resistance and rejection to change in the theory [4]. Although HPS contextualization has been proposed as a promising teaching approach for SRT [14,43], there are few articles that report on its implementation and evaluation.

Through analysis of popular physics textbooks in Argentina [38] and South Korea [10,44], it was found that HPS contextualization is poorly represented and that many textbooks contain flawed or oversimplified historical accounts. For example, it was shown by Gim [10] that many South Korean textbooks overstate the influence of the Michelson and Morley experiment, while Lorentz' and Pointcarè's pivotal influence in the development of SRT was underrepresented.

Arriassecq and Greca [45] reported a teaching proposal for SRT at the secondary school level that adopts a historically and epistemologically contextualized approach. A teaching sequence was designed to connect learning difficulties, identified by the authors in a previous (Spanish) study, to learning objectives related to the concepts of space, time, spacetime, observers, simultaneity, and measurement. The resulting teaching material encompassed a historical account of the origin of SRT, as well as epistemological issues concerning the role of experimentation in SRT, the originality of SRT and its influence on science, philosophy, and arts. Results of a pre-post test, analysis of student activities, field notes, and audio recordings ($N = 27$) indicated a positive effect on motivational aspects and student understanding of the concepts addressed.

### C. Multimedia support

Over the last three decades, multimedia applications, varying from interactive simulation tools to (immersive) virtual and augmented reality environments, have been shown to play an important role in students' motivation and attitude toward science [46]. In SRT education, the advantage of multimedia applications is twofold: first, they offer the possibility to visualize the world as seen while traveling at velocities near the speed of light, and second, they assist students in the construction, performance, and evaluation of TEs. In particular, some multimedia applications offer the possibility to view motion from different frames of reference, and to switch between them easily. This advantage over traditional pen-and-paper activities has the potential to support students in overcoming learning difficulties with regard to frames of reference. This section presents multimedia applications found in the selected articles divided into two categories: realistic virtual environments and visualization of TEs.

*1. Realistic virtual environments*

In order to bridge the gap between the abstract concepts of SRT and everyday experience, a number of relativistic virtual environments have been designed in which students can experience the effects of traveling at velocities near the speed of light.

McGrath *et al.* [47] reported the development of *Real Time Relativity*, a three-dimensional gamelike environment from a first-person perspective. By accelerating the player's vehicle to velocities near the speed of light, relativistic effects such as time dilation and length contraction, as well as relativistic optical effects, such as Doppler shift and headlight effect, become more clearly visible. During three-hour sessions, small groups of graduate students played *Real Time Relativity* in pairs. After familiarizing themselves with the interface and environment, students were asked to determine how length contraction appears in *Real Time Relativity* and to carry out a virtual experiment to quantitatively measure time dilation. Analysis of a multiple choice pre-post test and confidence logs [$(N) = 300$], showed that students described the topic as being less abstract after playing *Real Time Relativity* and answered concept questions more correctly.

Carr and Bossomaier [48] reported *Relativistic Asteroids* a two-dimensional third-person perspective game that reimagines the classic video arcade game "Asteroids." The objective of the game is to maneuver a spaceship trough an asteroid field while earning points by shooting at the asteroids. The speed of light can be set from infinity, corresponding to the nonrelativistic limit, to a lower value that transforms the kinematics of the game from classic to relativistic. In the relativistic gameplay, moving objects become contracted and colors change in accordance with the Doppler shift. Analysis of a survey containing both concept questions and evaluative questions ($N = 67$) showed that participants were more capable of answering concept questions correctly and that the game made the topic of SRT more relatable and entertaining.

Kortemeyer [49] reported *A Slower Speed of Light*, a three-dimensional serious game, where the player is able to move freely in a relativistic world seen from a first-person perspective. *A Slower Speed of Light* was built using *OpenRelativity* [50], an open-source toolkit to stimulate relativistic effects within the popular *Unity* game engine. The main objective of the game is to slow down the speed of light by collecting *orbs*. As the speed of light slows down, the kinematics and visual appearance of the environment become more relativistic, and it becomes more challenging to collect more *orbs*. The game is designed to follow a "flipped" teaching philosophy, where the player is confronted with relativistic effects first, and receives and explanation of SRT afterward. In a study by Croxton [51], community-posted gameplay videos ($N = 20$) were scanned for segments where educational objectives, attitude, and measures of engagement were addressed. While players and viewers were found to engage enthusiastically with the concepts of SRT, the objective of the game, namely, to make SRT less paradoxical and more intuitive, was not reached. It is suggested that this is due to the flipped pedagogy of the game.

A number of the reported realistic virtual environments have not yet been fully implemented and empirically evaluated. De Hosson *et al.* [52] developed an immersive three-dimensional environment that resembles a billiards table with billiard balls moving at velocities near the speed of light. By applying an impulse to the billiard balls, the viewer observes changes in the shape of the ball, apparent nonsimultaneity of the bounces and the aberration of light. The authors expect that this experience will help students to develop a stronger intuition on relativistic behavior. Sherin *et al.* [53] reported *Einstein's playground*, an immersive planetarium show that was built using the *OpenRelativity* toolkit [50]. In the planetarium show, everyday phenomena such as moving duck boats and synchronized fireworks were displayed at different values of the speed of light. At slower speeds of light, the consequences of the finite speed of light become more apparent, as the length of the boats become contracted, the fireworks are no longer simultaneous and a Doppler shift will begin to appear. By visualizing relativistic phenomena in immersive everyday scenarios, the authors aim to make SRT more apprehensible for a general audience.

*2. Visualization of thought experiments*

Rather than creating realistic visualizations of relativistic effects, some multimedia applications focus more on assisting students to perform and evaluate TEs.

Horwitz and Barowy [54] reported the development of *RelLab*, an exploratory simulation tool that allows students to model and simulate TEs. After placing objects on a two-dimensional grid and assigning velocities to each object, the user is presented with an animated representation of the objects' trajectories. Additionally, the user is able to "pin down" an object, producing a simulation of the trajectories as seen from that object's frame of reference. *RelLab* was embedded into a cyclic teaching sequence in which students were asked to predict, perform, and evaluate selected TEs using the simulation tool. The effect of this teaching sequence was evaluated by a diagnostic open-ended questionnaire given to first year ($N = 112$) and second year ($N = 22$) undergraduate physics students. After the teaching sequence, clinical interviews were conducted to identify any change in students' reasoning. It was concluded that the possibility to switch between frames of reference provides a way of supporting student reasoning that is unavailable through static event diagrams. Moreover, students who had worked with *RelLab* were more successful in answering open-ended concept questions compared to an earlier study by Villani and Pacca [19].

A number of the reported multimedia applications that aim at visualizing TEs have not yet been fully implemented and empirically evaluated. Belloni et al. [22] reported a series of Java applets that aim to familiarize students with the concepts of clock synchronization, relativity of simultaneity and time dilation and length contraction. These applets provide an animation of the propagation of light rays in a number of TEs, and the viewer is able to pause, accelerate, or reverse the animation. One of the applets address the apparent *barn-pole* paradox by animating the barn's and pole's worldlines in a spacetime diagram. Similar educational software has been reported by Moraru et al. [24] and Kashnikov et al. [26]. Underwood and Zhai [25] demonstrated a smartphone app that uses the device's GPS receiver to calculate the time dilation of a moving device as measured by a stationary observer. Because the app uses the device's GPS location, however, it can only be used to calculate time dilation as measured by an observer who is stationary with respect to the Earth's frame of reference. Hence, this app does not support the symmetry of time measurements by observers in different inertial frames of reference.

### D. Student activities

Besides multimedia supported learning activities, some researchers have studied the effects of "unplugged" student activities. The main motivation behind the design of student activities is that they enable students to give meaning to abstract concepts [55]. This section presents student activities identified in the selected articles.

Guisasola et al. [56] described the design of a teaching sequence for undergraduate engineering students that includes a visit to an exhibition on SRT in a science museum. During previsit and postvisit lessons, selected relativistic situations were analyzed and discussed, while the activities in the museum aimed to provide meaning and interest to the concepts discussed. Analysis of a pre-post visit questionnaire ($N = 35$) showed that the activities effectively increased students' understanding of SRT, especially regarding the invariance of the speed of light. Students' postvisit responses also contained significantly more scientific arguments and mentioned more scientific-technical applications compared to the previsit responses.

Yildiz [57] investigated the effect of a writing activity, in which preservice teachers were asked to write a summary of SRT for secondary education students. By comparing the results of an open-ended pre-post test of an experimental group ($N = 34$) with a control group ($N = 39$), it was found that students who had engaged in the writing activity were more capable of answering the questions correctly. Moreover, it was found that the writing activity enabled students to organize scientific thoughts in their own words and to remember them more easily.

Chiarello [58] presented a didactic board game called *Time Race*, which was developed to illustrate the effect of time dilation. As players move between connected nodes on the board, the running speed of their clocks is adjusted according to their velocity. A self assessment survey of the players ($N = 591$) indicated an increase of interest in and comprehension of the phenomenon of time dilation. Because the amount of time dilation is determined only with respect to the board game's frame of reference, the game does not reflect the equivalence of time measurements by observers in different inertial frames of reference.

Alvarez et al. [59] investigated the inclusion of peer instruction to address student conceptions in secondary education. After a brief lecture, a multiple-choice conceptual question related to the lecture was given to the class. If the percentage of correct answers was between 30% and 70%, a peer discussion followed, and the students could reenter their answer. The authors developed a pre-post test to investigate student conceptions, and administered it to the participants ($N = 25$), in addition to a questionnaire addressing students' beliefs and attitudes toward SRT. The results showed that peer instruction has a strong positive impact on students' beliefs and attitudes, and it has potential to address student conceptions.

### E. Synthesis

A variety of teaching approaches has been reported to help students overcome learning difficulties. Although there is much overlap, the reported teaching approaches aim at distinct learning objectives. The main learning objectives are (i) to foster conceptual understanding of SRT; (ii) to foster understanding of SRT as an example of the nature of science; (iii) to gain motivation, confidence, and attitude toward learning SRT and physics in general. Broadly speaking, conceptual approaches aim (at least) at the first objective, HPS-integrated approaches aim at the second objective, while multimedia supported approaches and student activities tend to focus more on the third objective.

In almost all of the teaching approaches, TEs are used as a means to express conceptual features of SRT. Using carefully designed instructional materials and learning activities, TEs can be used to elicit students' existing conceptions and beliefs and to evaluate them after instruction. To achieve this, students should be actively engaged in the process of constructing and performing TEs and evaluating their outcome.

The way in which the TEs are presented and performed, and the level of student engagement, varies considerably across different teaching approaches. Multimedia supported teaching approaches that assist students in performing TEs, such as *RelLab* [54], have been shown to enhance students' conceptual understanding of SRT. Realistic virtual environments and serious games, such as *A Slower Speed of Light* [51], on the other hand, have not yet proven to enhance conceptual understanding. Rather, by bridging the gap between the abstract consequences of SRT and students' everyday experience, realistic virtual environments were shown to be successful in advancing motivational and attitudinal objectives.

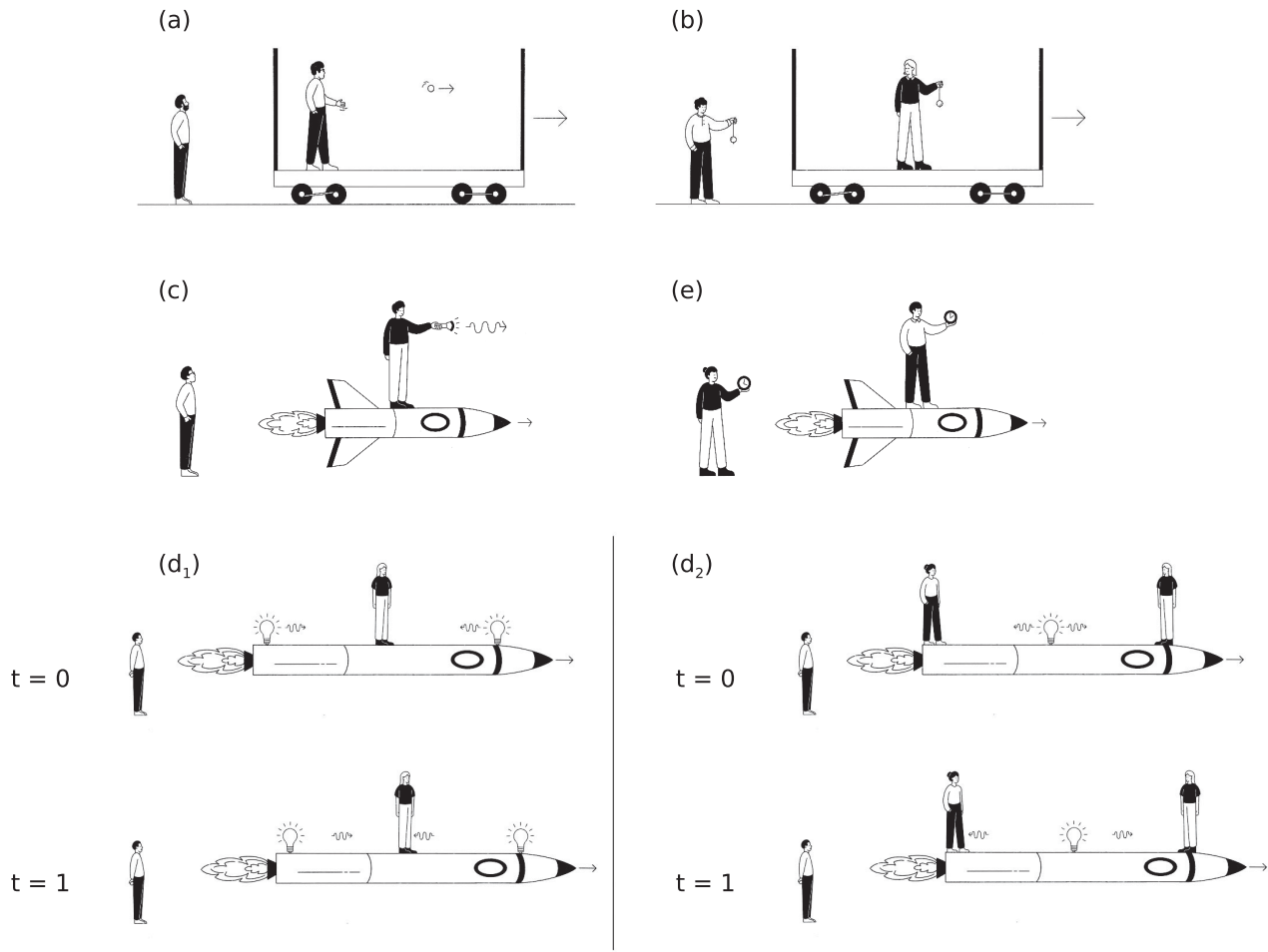

FIG. 3. Schematic illustrations of typical scenarios for TEs in the following categories: (a) Galilean transformation, (b) principle of relativity, (c) light postulate, ($d_1$) and ($d_2$) relativity of simultaneity, (e) time dilation, length contraction, and relativistic velocity addition. See the text for further explanation of the scenarios.

## V. RESEARCH TOOLS

This section addresses the third research question: "Which research tools have been used to probe students' conceptual understanding of SRT at the secondary and lower undergraduate level?" Ideally, learning outcomes of teaching approaches are measured by means of validated standard assessment tools, paving the way for quantitative comparison of teaching approaches. In the research field of SRT education, however, the development and evaluation of assessment tools is still at an early stage. One concept inventory could be identified in the selected literature [34], however there is a shortage of data on its versatility and reliability. Rather, the selected literature describes a variety of questionnaires, interview questions, and tasks that are designed to probe students' understanding of topics or skills related to a specific study. Typically, test items in these questionnaires lead students to perform and evaluate carefully crafted TEs scenarios. In this section, we present an analysis of the TE scenarios used as research tools in the selected literature and describe how they are used in the reported concept inventory. Finally, we discuss future directions for the development of research tools.

### A. Thought experiments

TEs are designed to express a particular concept or set of concepts in a context that is familiar to students. As a consequence, TEs that address the same concept, usually describe similar everyday scenarios. In this section, we describe the TE scenarios found in the selected literature grouped into five categories, corresponding roughly to the learning progression described in Sec. III: Galilean transformation, the principle of relativity, light postulate, relativity of simultaneity and time dilation, length contraction, and relativistic velocity addition. Schematic illustrations of typical scenarios for each category of TEs are shown in Fig. 3.

### 1. Galilean transformation

TEs in Galilean relativity involve measurements of distance, time intervals, velocity, and trajectories by observers in different inertial frames of reference [17,28,34,35,39,60]. In a typical TE, two inertial frames of reference are (implicitly) defined by introducing two observers, one often in rest with respect to the Earth and the other moving with respect to the Earth. The outcome of a measurement made by either one of the observers is given, and the student is asked to describe the motion as measured by an observer in the other frame of reference. As an example of a typical scenario, consider an observer throwing a ball in a train carriage, which is moving with respect to a platform, and a second observer that is in rest with respect to the platform, as illustrated in Fig. 3(a). According to Galilean transformation, the observers would disagree on the velocity of the ball and the distance travelled in a certain time interval. Other scenarios in this category include rowing a boat across a stream, passing cars on a freeway, or viewing a car chase from a helicopter.

Student conceptions described in Sec. III A were identified using this type of TE. Notably, Horwitz and Barowy [54] and Ramadas et al. [29] used this type of TE to investigate students' understanding of Galilean invariance of time intervals. This type of TE is widely used in multimedia supported teaching approaches, as described in Sec. IV C 2.

A variation of this type of TE, where one of the two frames of reference is noninertial, was used by Ramadas et al. [35] to investigate student conceptions of noninertial frames of reference, as described in Sec. III A 1.

### 2. Principle of relativity

The principle of relativity was first introduced by Galilei by a simple TE presently known as *Galileo's ship*. In his 1632 book [61], Galilei argues that it is impossible to inquire into a ship's movement relative to the Earth exclusively from natural phenomena observed from within the ship. From the outcome of this TE, Galilei concluded that the laws of motion must be equal in each inertial frame of reference. Many variants of this TE were found in the selected literature concerning the principle of relativity [32,34,39]. Typically, some physical experiment is described, such as the swinging of a pendulum, and students are asked whether the outcome of the experiment depends on whether it is carried out in uniform motion or in acceleration, as illustrated in Fig. 3(b).

This type of TE was used to investigate how students interpret the principle of relativity and whether they apply it spontaneously, as described in Sec. III B 1.

### 3. Light postulate

TEs that involve the light postulate require students to advance from classical kinematics to relativistic kinematics [6,19,34,48]. Typically, the one-way or two-way speed of light is measured in an inertial frame of reference and the student is asked to determine the speed of light as measured in another inertial frame of reference. Scenarios for this type of TE typically involve a carriage or rocket moving with respect to an observer and sending out or receiving a light signal, as illustrated in Fig. 3(c). These scenarios are similar to those used for TEs regarding Galilean transformation, as described above, the difference being that the propagation of light is considered rather than moving bodies.

These TEs require students to recognize that Galilean transformation fails to describe the propagation of light. Hence, they can be used to evaluate students' existing conceptions of light propagation, as described in Sec. III B 2.

### 4. Relativity of simultaneity

The concept of simultaneity is central to SRT, hence TEs regarding the relativity of simultaneity are well represented in the analyzed questionnaires [16,18,34,36,39]. In this type of TE, two spacelike separated events that are either simultaneous or nonsimultaneous in a certain inertial frame of reference are described, and the student is asked to assert the simultaneity or time sequence of the events in another inertial frame of reference. Two distinct scenarios were found for this type of TEs: (i) two spacelike separated flashes of light measured by an observer at rest with respect to the light sources, as illustrated in Fig. 3 ($d_1$); (ii) a single flash of light traveling toward two observers who receive the flashes at spacelike separated intervals, as illustrated in Fig. 3 ($d_2$). The first variant was used by Einstein in his famous 1916 TE [62], in which a train carriage is hit by two lightning strikes at its far ends.

These TEs are used to lead students to recognize that observers in different inertial frames of reference measure unique time intervals between two spacelike separated events. The learning difficulties described in Sec. III C 1, such as the tendency to associate the time of an event with the time at which an observer receives a light signal from that event, were investigated using this type of TE.

### 5. Time dilation, length contraction, and relativistic velocity addition

Relativistic effects only become apparent by comparative measurements by observers in different inertial frames of reference. TEs that express relativistic effects involve measurements of distance, time interval, or velocity in a certain inertial frame of reference, and the student is asked to compare the outcome with a measurement made by an observer in another inertial frame of reference [20,21,34,39,48]. Scenarios described in these TEs include measuring the length of a rod by means of two-way light signaling, or measuring the time interval between two events in a spaceship that is moving relative to an observer, as illustrated in Fig. 3(e). In this regard, these scenarios are similar to those reported for TEs regarding Galilean transformation and the light postulate, as described above, the difference being that these TEs require relativistic kinematics rather than classical kinematics. The famous "twin paradox" [63] also falls under this category, although resolving this TE actually requires switching between more than two inertial frames of reference.

TEs in this category are used to describe what the relativistic world looks like. Hence, these TEs are well represented in multimedia supported teaching approaches that focus on visualization, as described in Sec. IV C 2.

### B. Relativity concept inventory

Concept inventories, such as the popular Force Concept Inventory [64], are research-based tests that are used as research tools to measure learning outcomes of educational innovations or interventions. The Relativity Concept Inventory (RCI), aimed at measuring conceptual knowledge of SRT, was recently developed by Aslanides and Savage [34]. To our knowledge, the RCI is the only concept inventory for SRT currently reported.

The RCI measures conceptual knowledge of nine concepts, which were selected from relevant textbooks, physics education research literature, and feedback from discipline experts. The selected concepts are principle of relativity, light postulate, time dilation, length contraction, relativity of simultaneity, inertial frame of reference, velocity addition, causality, and mass-energy equivalence. Notably, causality is listed as one of the concepts, as it requires the invariance of time ordering of timelike separated events. The concepts are represented in 24 multiple-choice questions, each question describing a particular TE that expresses one of the concepts. The answers to the questions were designed to include distractors that represent common student conceptions, so that the test can be used to investigate the popularity of student conceptions.

The test was validated by an iterative process that combined expert feedback with student feedback on the clarity of the formulations. To test its potential to measure learning outcomes, the RCI was administered to a group of undergraduate students prior to instruction ($N = 70$), and after instruction ($N = 63$). The results of a statistical analysis of the responses suggest that the RCI may be too easy and, hence, insufficiently discriminating. To improve this, the authors recommend several revisions to the original questions. The results also show a significant gender difference.

### C. Future development of research tools

In order to investigate which of the teaching approaches offer the most promising strategy to overcome the reported learning difficulties, quantitative comparison of learning outcomes is needed. For this purpose, there is a need for a more thorough evaluation of assessment tools. The only validated research tool currently available is the RCI by Aslanides and Savage [34], although more data on student responses is required to determine its reliability and versatility, especially for use in secondary education.

Ideally, assessment tools should cover each step of the hypothetical learning progression described in Sec. III, by including at least one TE scenario from each of the five categories in our analysis. Currently, this criterion is only met by the RCI [34]. Questions about relativistic velocity addition are particularly underrepresented, with the RCI being the only research tool that includes a question on this topic.

As discussed in Sec. IV, three learning objectives can be identified in the reported teaching approaches: to foster conceptual understanding, to foster understanding of the history and philosophy of science, and to gain motivation and confidence. Future assessment tools should match the learning objective(s) of the teaching approach under evaluation. Teaching approaches that aim at conceptual understanding are mostly evaluated by means of TEs, while teaching approaches that focus on motivational and attitudinal objectives are mostly evaluated by questionnaires that inquire into the gained motivation and confidence. Although a better understanding of the conceptual and epistemological leap between classical and modern physics is regarded as one of the main learning objectives of SRT, no research tools have yet been reported that inquire into students' understanding of the role of SRT in the history and philosophy of science. Such assessment tools would inquire more deeply into students' existing conceptions of space, time, and motion as well as students' views on theory development in science.

## VI. CONCLUSION

This review provides an overview of the current body of research on learning difficulties, teaching approaches, and research tools in SRT education at the secondary and lower undergraduate level. We first present the main conclusions from each section, and subsequently discuss implications of our review for researchers and teachers.

The reviewed articles show that students experience learning difficulties with the use of frames of reference, the postulates of SRT, and relativistic effects. These learning difficulties occur at the level of secondary, undergraduate, and preservice teacher education. Understanding relativistic effects presupposes a solid understanding of the postulates, which in turn requires the ability to switch between frames of reference. Therefore, some researchers have argued that a stronger emphasis on Galilean transformation, perhaps at an earlier age, has the potential to help students overcome some of the reported learning difficulties. To foster meaningful understanding of relativistic effects, however, equal attention should be given to the invariance of the speed of light and its incompatibility with Galilean transformation.

A variety of teaching sequences, instructional materials, and learning activities has been reported to help students overcome learning difficulties. From the teaching approaches reported, three distinct learning objectives can be delineated: (i) to foster conceptual understanding; (ii) to foster understanding of the history and philosophy of science; (iii) to gain motivation, confidence, and attitude toward SRT and physics in general. Conceptual teaching approaches mainly aim at the first objective, HPS-contextualized teaching approaches aim at the second objective, while multimedia supported teaching approaches tend to focus more on the third objective.

In order to quantitatively compare learning outcomes of various teaching approaches, further development and evaluation of assessment tools is required. Typically, assessment tools lead students to perform and evaluate carefully crafted TE scenarios. From the selected literature, five groups of TE scenarios could be identified, focusing, respectively, on Galilean transformation, principle of relativity, light postulate, relativity of simultaneity and time dilation, length contraction, and relativistic velocity addition. Currently, the Relativity Concept Inventory by Aslanides and Savage [34] is the only validated research tool available, although more data are needed on its reliability and versatility.

### A. Implications for researchers

Because SRT education is represented by a relatively young field of research, especially at the level of secondary education, many gaps of knowledge presently exist. In this section, we highlight three topics that require further empirical inquiry.

One of the main objectives of SRT education is to introduce students to modern physics, and its radical departure from classical physics. A few studies on historical and philosophical contextualization have reported promising results [14,43,45], although they have not been thoroughly empirically evaluated. Further research is needed to explore how key issues from the philosophy of space and time as well as the role of SRT in the history of physics can be implemented in an introductory teaching sequence.

Difficulties in understanding and applying relativistic effects can often be traced back to difficulties in understanding the postulates on which they are based, as well as prerequisite knowledge about the use of frames of reference. This problem underscores the potential power of TEs, as they allow students to engage actively with underlying principles and assumptions, and to derive relativistic effects from them. In many of the multimedia supported learning tools, including realistic virtual realities, the possibility to actively perform TEs is obscured, as the outcomes of TEs are usually pre-programmed. More research is needed to investigate how realistic virtual environments can support an active role for students in predicting, performing, and evaluating TEs.

It is notable that some of the most robust student conceptions are reflected by popular TE scenarios. For example, TE scenarios that describe measurements with respect to the Earth's frame of reference may support the special status that students often ascribe to the Earth's frame of reference, which is at odds with the principle of relativity. It remains to be investigated whether TE scenarios that take place in an otherwise empty space would remove this tendency toward a preferred frame of reference. Similarly, TEs that express the concept of simultaneity seemingly tend to favor the association of the time of an event with the time at which an observer receives a light signal from that event. Students may interpret this category of TE as falsely suggesting that events are simultaneous when an observer receives light signals instantaneously. More research is needed to investigate whether a stronger focus on the notion of time measurement by means of synchronized clocks would help to resolve this learning difficulty. Finally, it was suggested by Selçuk [20] that teachers' word choice may also play an important role in students' understanding of SRT. According to Selçuk, teachers should be careful when using words such as "see" or "appear," as they might induce the impression that the observed phenomenon is merely a matter of perception. To our knowledge, no research has yet been performed on the effects of linguistic issues.

### B. Implications for teachers

Learning SRT presents a notorious challenge for students at all levels of education, as evidenced by the variety and robustness of the reviewed learning difficulties. Many of these learning difficulties are rooted in difficulties in basic understanding of the use of frames of reference and the role of observers and events. Consequently, a stronger emphasis on switching between frames of reference may be a fruitful first step to resolving some of the reported learning difficulties.

TEs are a helpful tool to bridge the gap between the abstract concepts of SRT and everyday experience. As argued by Scherr [16,36] and Reiner and Burko [37], the use of TEs is most fruitful when they are carried out by the students themselves, rather than presented in a textbook or by the teacher. Bringing interactive student activities, such as multimedia applications, into the classroom increases student engagement in the process of constructing and evaluating TEs.

Finally, when describing the motion of a body, it is useful to refer to a specific frame of reference, even if the motion is described with respect to the Earth's frame of reference. In particular, caution should be made when using words such as stationary, standing still, or moving without referring to a frame of reference, as this choice of words naturally supports the conception of a preferred frame of reference. Similarly, caution should be made when describing observers as "seeing" an event, as this could support the conception that the time ordering of spacelike-separated events is dependent on the position of the observer.

## ACKNOWLEDGMENTS

This work was supported by The Dutch Research Council (NWO) under Grant No. 023.013.003.

[1] J. Osborne, S. Collins, M. Ratcliffe, R. Millar, and R. Duschl, What ideas-about-science should be taught in school science? A delphi study of the expert community, J. Res. Sci. Teach. **40,** 692 (2003).

[2] P. Kind and J. Osborne, Styles of scientific reasoning: A cultural rationale for science education?, Sci. Educ. **101,** 8 (2017).

[3] J. K. Cosgrove, Minkowski's "Space and Time", in *Relativity without Spacetime* (Springer International Publishing, Cham, 2018), pp. 11–33.

[4] A. Villani and S. M. Arruda, Special theory of relativity, conceptual change and history of science, Sci. Educ. **7,** 85 (1998).

[5] M. Kersting, E. K. Henriksen, M. V. Bøe, and C. Angell, General relativity in upper secondary school: Design and evaluation of an online learning environment using the model of educational reconstruction, Phys. Rev. Phys. Educ. Res. **14,** 010130 (2018).

[6] F. Kamphorst, M. J. Vollebregt, E. R. Savelsbergh, and W. R. van Joolingen, Students' preinstructional reasoning with the speed of light in relativistic situations, Phys. Rev. Phys. Educ. Res. **15,** 020123 (2019).

[7] C. Zahn and U. Kraus, Sector models—A toolkit for teaching general relativity: I. Curved spaces and spacetimes, Eur. J. Phys. **35,** 055020 (2014).

[8] M. Otero, M. Arlego, and E. Munoz, Relativity of simultaneity in secondary school: An analysis based on the Theory of the Conceptual Fields, J. Phys. Conf. Ser. **1287,** 012002 (2019).

[9] T. Kaur, D. Blair, J. Moschilla, W. Stannard, and M. Zadnik, Teaching Einsteinian physics at schools: Part 1, models and analogies for relativity, Phys. Educ. **52,** 1 (2017).

[10] J. Gim, Special theory of relativity in South Korean high school textbooks and new teaching guidelines, Sci. Educ. **25,** 575 (2016).

[11] T. Kaur, D. Blair, W. Stannard, D. Treagust, G. Venville, M. Zadnik, W. Mathews, and D. Perks, Determining the intelligibility of Einsteinian concepts with middle school students, Res. Sci. Educ. **50,** 2505 (2020).

[12] R. Choudhary, U. Kraus, M. Kersting, D. Blair, C. Zahn, M. Zadnik, and R. Meagher, Einsteinian Physics in the classroom: Integrating physical and digital learning resources in the context of an international research collaboration, Phys. Educator **01,** 1950016 (2019).

[13] K. Dimitriadi and K. Halkia, Secondary students' understanding of basic ideas of special relativity, Int. J. Sci. Educ. **34,** 2565 (2012).

[14] O. Levrini, The role of history and philosophy in research on teaching and learning of relativity, in *International Handbook of Research in History, Philosophy and Science Teaching*, edited by M. R. Matthews (Springer, New York, 2014), Chap. 6, pp. 157–181.

[15] R. Duit, D. F. Treagust, and A. Widodo, Teaching science for conceptual change: Theory and practice, in *International Handbook of Research on Conceptual Change*, edited by S. Vosniadou (Taylor and Francis Inc., London, 2013), pp. 487–503.

[16] R. E. Scherr, P. S. Shaffer, and S. Vokos, Student understanding of time in special relativity: Simultaneity and reference frames, Am. J. Phys. **69,** S24 (2001).

[17] E. Saltiel and J. L. Malgrange, 'Spontaneous' ways of reasoning in elementary kinematics, Eur. J. Phys. **1,** 73 (1980).

[18] A. Velentzas and K. Halkia, The use of thought experiments in teaching physics to upper secondary-level students: Two examples from the theory of relativity, Int. J. Sci. Educ. **35,** 3026 (2013).

[19] A. Villani and J. L. Pacca, Students' spontaneous ideas about the speed of light, Int. J. Sci. Educ. **9,** 55 (1987).

[20] G. S. Selçuk, Addressing pre-service teachers' understandings and difficulties with some core concepts in the special theory of relativity, Eur. J. Phys. **32,** 1 (2011).

[21] Z. Tanel, Student difficulties in solving problems concerning special relativity and possible reasons for these difficulties, J. Baltic Sci. Educ. **13,** 573 (2014).

[22] M. Belloni, W. Christian, and M. H. Dancy, Teaching special relativity using Physlets®, Phys. Teach. **42,** 284 (2004).

[23] R. Barbier, S. Fleck, S. Perriès, and C. Ray, Integration of information and communication technologies in special relativity teaching, Eur. J. Phys. **26,** S13 (2005).

[24] S. Moraru, I. Stoica, and F. F. Popescu, Educational software applied in teaching and assessing physics in high schools, Rom. Rep. Phys. **63,** 577 (2011).

[25] B. Underwood and Y. Zhai, Moving phones tick slower: Creating an android app to demonstrate time dilation, Phys. Teach. **54,** 277 (2016).

[26] A. V. Kashnikov, A. A. Gusmanova, and E. O. Kiktenko, Demonstration of special relativity effects with specialized software, J. Phys. Conf. Ser. **1348,** 012092 (2019).

[27] A. Einstein, On the electrodynamics of moving bodies, Ann. Phys. (Berlin) **322,** 187 (1905).

[28] S. Panse, J. Ramadas, and A. Kumar, Alternative conceptions in Galilean relativity: Frames of reference, Int. J. Sci. Educ. **16,** 63 (1994).

[29] J. Ramadas, S. Barve, and A. Kumar, Alternative conceptions in Galilean relativity: Inertial and non-inertial observers, Int. J. Sci. Educ. 18, 615 (1996).

[30] C. De Hosson, I. Kermen, and E. Parizot, Exploring students' understanding of reference frames and time in Galilean and special relativity, Eur. J. Phys. 31, 1527 (2010).

[31] M. Pietrocola and A. Zylbersztajn, The use of the principle of relativity in the interpretation of phenomena by undergraduate physics students, Int. J. Sci. Educ. 21, 261 (1999).

[32] A. Bandyopadhyay, Students' ideas of the meaning of the relativity principle, Eur. J. Phys. 30, 1239 (2009).

[33] O. Ozcan, Examining the students' understanding level towards the concepts of special theory of relativity, Problems Educ. 21st Century 75, 263 (2017).

[34] J. S. Aslanides and C. M. Savage, Relativity concept inventory: Development, analysis, and results, Phys. Rev. ST Phys. Educ. Res. 9, 010118 (2013).

[35] J. Ramadas, S. Barve, and A. Kumar, Alternative conceptions in galilean relativity: Distance, time, energy and laws, Int. J. Sci. Educ. 18, 463 (1996).

[36] R. E. Scherr, P. S. Shaffer, and S. Vokos, The challenge of changing deeply held student beliefs about the relativity of simultaneity, Am. J. Phys. 70, 1238 (2002).

[37] M. Reiner and L. Burko, On the limitations of thought experiments in physics and the consequences for physics education, Sci. Educ. 12, 365 (2003).

[38] I. Arriassecq and I. M. Greca, Approaches to the teaching of special relativity theory in high school and university textbooks of Argentina, Sci. Educ. 16, 65 (2007).

[39] M. R. Otero, M. Arlego, and F. Prodanoff, Teaching the basic concepts of the Special Relativity in the secondary school in the framework of the Theory of Conceptual Fields of Vergnaud, Nuovo Cimento della Societa Italiana di Fisica C 38, 108 (2015).

[40] I. Galili and A. Hazan, Experts' views on using history and philosophy of science in the practice of physics instruction, Sci. Educ. 10, 345 (2001).

[41] D. Höttecke and C. C. Silva, Why implementing history and philosophy in school science education is a challenge: An analysis of obstacles, Sci. Educ. 20, 293 (2011).

[42] M. R. Matthews, Science Teaching: The Role of History and Philosophy of Science (Routledge, New York, 1994).

[43] O. Levrini, The substantivalist view of spacetime proposed by Minkowski and its educational implications, Sci. Educ. 11, 601 (2002).

[44] H. Kim and G. Lee, Reflection and outlook on special relativity education from the perspective of Einstein: Focusing on research papers published in Korea, J. Korean Phys. Soc. 73, 422 (2018).

[45] I. Arriassecq and I. M. Greca, A teaching-learning sequence for the special relativity theory at high school level historically and epistemologically contextualized, Sci. Educ. 21, 827 (2012).

[46] N. Rutten, W. R. Van Joolingen, and J. T. Van Der Veen, The learning effects of computer simulations in science education, Comput. Educ. 58, 136 (2012).

[47] D. McGrath, M. Wegener, T. J. McIntyre, C. Savage, and M. Williamson, Student experiences of virtual reality: A case study in learning special relativity, Am. J. Phys. 78, 862 (2010).

[48] D. Carr and T. Bossomaier, Relativity in a rock field: A study of physics learning with a computer game, Australas. J. Educ. Tech. 27, 1042 (2011).

[49] G. Kortemeyer, Game development for teaching physics, J. Phys. Conf. Ser. 1286, 012048 (2019).

[50] Z. W. Sherin, R. Cheu, P. Tan, and G. Kortemeyer, Visualizing relativity: The OpenRelativity project, Am. J. Phys. 84, 369 (2016).

[51] D. Croxton and G. Kortemeyer, Informal physics learning from video games: A case study using gameplay videos, Phys. Educ. 53, 015012 (2018).

[52] C. de Hosson, K. Isabelle, M. Clémént, P. Etienne, D. Tony, and V. Jean-Marc, Learning scenarios for a 3D virtual environment: The case of special relativity, in Springer Proceedings in Physics (Springer, New York, 2014), Vol. 145, pp. 377–383.

[53] Z. Sherin, P. Tan, H. Fairweather, and G. Kortemeyer, Einstein's playground: An interactive planetarium show on special relativity, Phys. Teach. 55, 550 (2017).

[54] P. Horwitz and B. Barowy, Designing and using open-ended software to promote conceptual change, J. Sci. Educ. Technol. 3, 161 (1994).

[55] E. Kocevar-Weidinger, Beyond active learning: A constructivist approach to learning, Reference Services Rev. 32, 2 (2004).

[56] J. Guisasola, J. Solbes, J. Barragues, M. Morentin, and A. Moreno, Students' understanding of the special theory of relativity and design for a guided visit to a science museum, Int. J. Sci. Educ. 31, 2085 (2009).

[57] A. Yildiz, Prospective teachers' comprehension levels of special relativity theory and the effect of writing for learning on achievement, Australian J. Teacher Educ. 37, 15 (2012).

[58] F. Chiarello, Board games to learn complex scientific concepts and the "Photonics Games" competition, in Proceedings of the European Conference on Games-based Learning (Academic Conferences International Limited, Oxford, 2015), Vol. 2015, pp. 774–779.

[59] M. Alvarez-Alvarado, C. Mora, and C. Cevallos-Reyes, Peer Instruction to address alternative conceptions in Einstein's special relativity, Revista Brasileira de Ensino de Fisica 41, 4 (2019).

[60] P. Horwitz, E. F. Taylor, and W. Barowy, Teaching special relativity with a computer, Comput. Phys. 8, 92 (1994).

[61] G. Galilei, A. Einstein, and S. Drake, Dialogue Concerning the Two Chief World Systems (University of California Press, Berkeley, CA, 2020).

[62] A. Einstein and R. W. Lawson, Relativity: The Special and the General Theory (Taylor and Francis, London, 2012), pp. 1–118.

[63] P. Pesic, Einstein and the twin paradox, Eur. J. Phys. 24, 585 (2003).

[64] D. Hestenes, M. Wells, and G. Swackhamer, Force Concept Inventory, Phys. Teach. 30, 141 (1992).